\documentclass[multicol,aps,pre,epsf,amsmath,citesort,twocolumn]{revtex4}
\usepackage{graphicx,amsfonts,amsmath}

\def \BEA {\begin{eqnarray}}
\def \EEA {\end{eqnarray}}
\def \BC {\begin{cases}}
\def \EC {\end{cases}}

\def \wt {\tilde{\omega}}
\def \Tt {\tilde{T}}
\def \tt {\tilde{t}}
\def \w {\omega}

\def \a {\alpha}
\def \b {\beta}
\def \g {\gamma}
\def \at {\tilde{a}}
\def \af {\hat{a}}
\def \la {\langle}
\def \ra {\rangle}
\def \D {\Delta}

\def \i {\imath}
\def \eqN {\overset{N}{=}}
\def \eq1 {\overset{1}{=}}

\def \eb {\bar{e}}

\def \T {\theta}
\def \tinf {\theta_{\infty}}

\begin{document}

\title{Interactions of renormalized waves in thermalized Fermi-Pasta-Ulam chains}
%
\author{Boris Gershgorin$^1$, Yuri V. Lvov$^1$ and David Cai$^2$, \\ \ \\
$^1$ \textit{Department of Mathematical Sciences, Rensselaer
Polytechnic
Institute, Troy, NY 12180}\\
$^2$ \textit{Courant Institute of Mathematical Sciences,
        New York University, New York, NY 10012}\\
}
\begin{abstract}
The dispersive interacting waves in Fermi-Pasta-Ulam (FPU) chains of
particles in \textit{thermal equilibrium} are studied from both
statistical and wave resonance perspectives. It is shown that, even
in a strongly nonlinear regime, the chain in thermal equilibrium can
be effectively described by a system of weakly interacting
\textit{renormalized} nonlinear waves that possess (i) the
Rayleigh-Jeans distribution and (ii) zero correlations between
waves, just as noninteracting free waves would. This renormalization
is achieved through a set of canonical transformations. The
renormalized linear dispersion of these renormalized waves is
obtained and shown to be in excellent agreement with numerical
experiments. Moreover, a dynamical interpretation of the
renormalization of the dispersion relation is provided via a
self-consistency, mean-field argument. It turns out that this
renormalization arises mainly from the trivial resonant wave
interactions, i.e., interactions with no momentum exchange.
Furthermore, using a multiple time-scale, statistical averaging
method, we show that the interactions of near-resonant waves give
rise to the broadening of the resonance peaks in the frequency
spectrum of renormalized modes. The theoretical prediction for the
resonance width for the thermalized $\beta$-FPU chain is found to be
in very good agreement with its numerically measured value.
\end{abstract}
\maketitle
\section{Introduction}
The study of discrete one-dimensional chains of particles with the
nearest-neighbor interactions provides insight to the dynamics of
various physical and biological systems, such as crystals, wave
systems, and biopolymers~\cite{fpu50, Peyrard, Toda}. In the thermal
equilibrium state, such nonlinear chains can be described by the
canonical Gibbs measure~\cite{LL5} with the Hamiltonian \BEA
H=\sum_j
\frac{p_j^2}{2}+\frac{(q_j-q_{j+1})^2}{2}+V(q_j-q_{j+1}),\label{intro_Hpq}
\EEA where $p_j$ and $q_j$ are the momentum and the displacement
from the equilibrium position of the $j$-th particle, respectively,
$V(q_j-q_{j+1})$ is the anharmonic part of the potential, and the
mass of each particle and the linear spring constant are scaled to
unity. In this article, we only consider the potentials of the
\textit{restoring} type, i.e., the potentials for which the Gibbs
measure exists. In order to study interactions of waves in such
systems, one usually introduces the canonical complex normal
variables $a_k$ via \BEA
a_k=\frac{P_k-\i\w_kQ_k}{\sqrt{2\w_k}},\label{intro_a} \EEA where
$P_k$ and $Q_k$ are the Fourier transforms of $p_j$ and $q_j$,
respectively, and $\w_k=2\sin(\pi k/N)$ is the linear dispersion
relation of the waves represented by $a_k$. In terms of the $a_k$,
the Hamiltonian~(\ref{intro_Hpq}) becomes \BEA
H=\sum\w_k|a_k|^2+V(a)\label{Ham_a}, \EEA where $V(a)$ is the
combination of various products of $a_k$ and $a_k^*$ corresponding
to various wave-wave interactions. If the potential in
Eq.~(\ref{intro_Hpq}) is harmonic, i.e., $V\equiv 0$, then $a_k$
correspond to ideal, \textit{free} waves, which have no energy
exchanges among different $k$ modes. In thermal equilibrium, the
Boltzmann distribution $ \exp(-\theta^{-1}\sum\w_k|a_k|^2) $ with
temperature $\T$, gives rise to the following properties of free
waves \BEA
\la a_k^*a_l\ra&=&n_k\delta^k_l,\label{intro_asa}\\
\la a_ka_l\ra&=&0\label{intro_aa}, \EEA for any $k$ and $l$, where
$n_k\equiv\la|a_k|^2\ra=\theta/\w_k$ is the power spectrum. If the
anharmonic part of the potential is sufficiently weak, then
corresponding waves $a_k$ remain almost free, and
Eqs.~(\ref{intro_asa}) and~(\ref{intro_aa}) would be approximately
satisfied in the weakly nonlinear regime. However, when the
nonlinearity becomes stronger, waves $a_k$ become strongly
correlated, and, in general, the correlations between waves
[Eq.~(\ref{intro_aa})] no longer vanish. In particular, $\la
a_ka_{N-k}\ra\neq 0$, as will be shown below. Naturally, the
question arises: can the strongly nonlinear system in thermal
equilibrium still be viewed as a system of almost free waves in some
statistical sense? In this article, we address this question with an
affirmative answer: it turns out that the system~(\ref{intro_Hpq})
can be described by a complete set of \textit{renormalized}
canonical variables $\at_k$, which still possess the wave properties
given by Eqs.~(\ref{intro_asa}) and~(\ref{intro_aa}) with a
renormalized linear dispersion. The waves that correspond to these
new variables $\at_k$ will be referred to as \textit{renormalized}
waves. Since these renormalized waves possess the equilibrium
Rayleigh-Jeans distribution~\cite{ZLF} and vanishing correlations
between waves, they resemble free, non-interacting waves, and can be
viewed as statistical normal modes. Furthermore, it will be
demonstrated that the renormalized linear dispersion for these
renormalized waves has the form $\wt_k=\eta(k)\w_k$, where $\eta(k)$
is the linear frequency renormalization factor, and is
\textit{independent} of $k$ as a consequence of the Gibbs measure.

In our method, the construction of the renormalized variables
$\at_k$ does not depend on a particular form or strength of the
anharmonic potential, as long as it is of the restoring type with
only the nearest neighbor interactions, as in Eq.~(\ref{intro_Hpq}).
Therefore, our approach is non-perturbative and can be applied to a
large class of systems with strong nonlinearity. However, in this
article, we will focus on the $\beta$-FPU chain to illustrate the
theoretical framework of the renormalized waves. We will verify that
$\at_k$ effectively constitute normal modes for the $\beta$-FPU
chain in thermal equilibrium by showing that (i) the theoretically
obtained renormalized linear dispersion relationship is in excellent
agreement with its dynamical manifestation in our numerical
simulation, and (ii) the equilibrium distribution of $\at_k$ is
still a Rayleigh-Jeans distribution and $\at_k$'s are uncorrelated.
Note that similar expressions for the renormalization factor $\eta$
have been previously discussed in the framework of an approximate
virial theorem~\cite{Alabiso} or effective long wave dynamics via
the Zwanzig-Mori projection~\cite{Lepri_renorm}. However, in our
theory, the exact formula for the renormalization factor is derived
from a \textit{precise} mathematical construction of statistical
normal modes, and is valid for all wave modes $k$ --- no longer
restricted to long waves.

Next, we address how renormalization arises from the dynamical wave
interaction in the $\beta$-FPU chain.
We will show that the $\beta$-FPU chain can be \textit{effectively}
described as a four-wave interacting Hamiltonian system of the
renormalized resonant waves $\at_k$. We will study the resonance
structure of the $\beta$-FPU chain and find that most of the exact
resonant interactions are \textit{trivial}, i.e., the interactions
with no momentum exchange among different wave modes. In what
follows, the renormalization of the linear dispersion will be
explained as a collective effect of these trivial resonant
interactions of the renormalized waves $\at_k$. We will use a
self-consistency argument to find an approximation, $\eta_{sc}$, of
the renormalization factor $\eta$. As will be seen below, the
self-consistency argument essentially is of a mean-field type, i.e.,
the renormalization arises from the scattering of a wave by a
mean-background of waves in thermal equilibrium via trivial resonant
interactions. We note that our self-consistency, mean-field argument
is not limited to the weak nonlinearity. Very good agreement of the
renormalization factor $\eta$ and its dynamical approximation
$\eta_{sc}$ --- for weakly as well as strongly nonlinear waves ---
confirms that the renormalization is, indeed, \textit{a direct
consequence} of the trivial resonances.

We will further study the properties of these renormalized waves by
investigating how long these waves are coherent, i.e., what their
frequency widths are. Therefore, we consider \textit{near-resonant}
interactions of the renormalized waves $\at_k$, i.e., interactions
that occur in the vicinity of the resonance manifold, since most of
the exact resonant interactions are trivial, i.e., with no momentum
exchanges, and they, cannot effectively redistribute energy among
the wave modes.

We will demonstrate that near-resonant interactions of the
renormalized waves $\at_k$ provide a mechanism for effective energy
exchanges among different wave modes. Taking into account the
near-resonant interactions, we will study analytically the frequency
peak broadening of the renormalized waves $\at_k$ by employing a
multiple time-scale, statistical averaging method. Here, we will
arrive at a theoretical prediction of the spatiotemporal spectrum
$|\af_k(\w)|^2$, where $\af_k(\w)$ is the Fourier transform of the
normal variable $\at_k(t)$, and $\w$ is the frequency. The predicted
width of frequency peaks is found to be in good agreement with its
numerically measured values.

In addition, for a finite $\beta$-FPU chain, we will mention the
consequence, to the correlation times of waves, of the momentum
exchanges that cross over the first Brillouin zone. This process is
known as the umklapp scattering in the setting of phonon
scattering~\cite{Umklapp}. Note that, in the previous
studies~\cite{Kramer} of the FPU chain from the wave turbulence
point of view, the effects arising from the finite  nature of the
chain were not taken into account, i.e., only the limiting case of
$N\rightarrow\infty$, where $N$ is the system size, was considered.

The article is organized as follows. In
Section~\ref{sect_Hformalism}, we discuss a chain of particles with
the nearest-neighbor nonlinear interactions. We demonstrate how to
describe a strongly nonlinear system as a system of waves that
resemble free waves in terms of the power spectrum and vanishing
correlations between waves. We show how to construct the
corresponding renormalized variables with the renormalized linear
dispersion. In Section~\ref{sect_FPU}, we rewrite the $\beta$-FPU
chain as an interacting four-wave Hamiltonian system. We study the
dynamics of the chain numerically and find excellent agreement
between the renormalized dispersion, obtained analytically and
numerically. In Section~\ref{sect_Dispersion}, we describe the
resonance manifold analytically and illustrate its controlling role
in long-time averaged dynamics using numerical simulation. In
Section~\ref{sect_Renormalization}, we derive an approximation for
the renormalization factor for the linear dispersion using a
self-consistency condition. In Section~\ref{sect_Width}, we study
the broadening effect of frequency peaks  and predict analytically
the form of the spatiotemporal spectrum for the $\beta$-FPU chain.
We provide the comparison of our prediction with the numerical
experiment. We present the conclusions in
Section~\ref{sect_Conclusions}.
\section{Renormalized waves}
\label{sect_Hformalism} Consider a chain of particles coupled via
nonlinear springs. Suppose the total number of particles is $N$ and
the momentum and displacement from the equilibrium position of the
$j$-th particle are $p_j$ and $q_j$, respectively. If only the
nearest-neighbor interactions are present, then the chain can be
described by the Hamiltonian \BEA H=H_2+V,\label{Ham_pq} \EEA where
the quadratic part of the Hamiltonian takes the form \BEA
H_2=\frac{1}{2}\sum_{j=1}^N{p_j^2}+(q_j-q_{j+1})^2,\label{H2} \EEA
and the anharmonic potential $V$ is the function of the relative
displacement $q_j-q_{j+1}$. Here periodic boundary conditions
$q_{N+1}\equiv q_1$ and $p_{N+1}\equiv p_1$ are imposed. Since the
total momentum of the system is conserved, it can be set to zero.

In order to study the distribution of energy among the wave modes,
we transform the Hamiltonian to the Fourier variables $Q_k$, $P_k$
via \BEA \BC
Q_k=\displaystyle{\frac{1}{\sqrt{N}}}\sum_{j=0}^{N-1}q_je^{\frac{2\pi \i kj}{N}},\\
P_k=\displaystyle{\frac{1}{\sqrt{N}}}\sum_{j=0}^{N-1}p_je^{\frac{2\pi
\i kj}{N}}. \EC\label{PQ} \EEA This transformation is
canonical~\cite{LL1,Licht} and the Hamiltonian~(\ref{Ham_pq})
becomes \BEA
H&=&\frac{1}{2}\sum_{k=1}^{N-1}|P_k|^2+\omega_k^2|Q_k|^2+V(Q),\label{Ham_PQ}
\EEA where $\w_k=2\sin(\pi k/N)$ is the linear dispersion relation.
Note that, throughout the paper, for the simplicity of notation, we
denote the periodic wave number space by the set of integers in the
range $[0,N-1]$, i.e., we drop the conventional factor, $2\pi/N$.
The zeroth mode vanishes due to the fact that the total momentum is
zero.

If the system~(\ref{Ham_PQ}) is in thermal equilibrium, then the
canonical Gibbs measure, with the corresponding partition function
\BEA Z=\int_{-\infty}^{\infty}e^{-H(p,q)/\theta}dpdq,\label{Zpq}
\EEA with the temperature $\theta$, can be used to describe the
statistical behavior of the system. We consider the systems with the
anharmonic potential of the restoring type, i.e., the potential for
which the integral in Eq.~(\ref{Zpq}) converges. It can be easily
shown that for system~(\ref{Ham_PQ}) the average kinetic energy
$K_k$ of each mode is independent of the wave number \BEA \la
K_k\ra=\la K_l\ra,\label{Kkind} \EEA where $k$ and $l$ are wave
numbers, $K_k\equiv|P_k|^2/2$, and $\la\dots\ra$ denotes averaging
over the Gibbs measure. Similarly, the average quadratic potential
$U_k$ of each mode is independent of the wave number \BEA \la
U_k\ra=\la U_l\ra,\label{Qkind} \EEA where
$U_k\equiv\w_k^2|Q_k|^2/2$.

If the nonlinear interactions are weak, then it is convenient to
further transform the Hamiltonian~(\ref{Ham_PQ}) to the complex
normal variables defined by Eq.~(\ref{intro_a}). This transformation
is canonical, i.e., the dynamical equation of motion becomes \BEA
\i\dot{a}_k=\frac{\partial H}{\partial a_k^*}.\label{a_canonical}
\EEA In terms of these normal variables, the
Hamiltonian~(\ref{Ham_PQ}) takes the form~(\ref{Ham_a}). For the
system of noninteracting waves, i.e.,
$H=\sum_{k=1}^{N-1}\w_k|a_k|^2$, we obtain a standard virial theorem
in the form \BEA \la K_k\ra\vline_{V=0}=\la
U_k\ra\vline_{V=0}.\label{free} \EEA As a consequence of this virial
theorem, we have the properties of free waves, which were already
mentioned above [Eqs.~(\ref{intro_asa}) and~(\ref{intro_aa})], i.e.,
\BEA
\la a_k^*a_l\ra&=&\frac{1}{2\w_k}(\la|P_k|^2\ra+\w_k^2\la|Q_k|^2\ra)\delta^k_l=\frac{\theta}{\w_k}\delta^k_l,\label{asa}\\
\la
a_ka_l\ra&=&\frac{1}{2\w_k}(\la|P_k|^2\ra-\w_k^2\la|Q_k|^2\ra)\delta^{k+l}_N=0,\label{aa}
\EEA for all wave numbers $k$ and $l$. Note that
equation~(\ref{asa}) gives the classical Rayleigh-Jeans distribution
for the power spectrum of free waves~\cite{ZLF} \BEA
n_k=\frac{\T}{\w_k}.\label{PSa} \EEA However, if the nonlinearity is
present, the waves $a_k$ and $a_{N-k}$ become correlated, i.e., \BEA
\la
a_ka_{N-k}\ra=\frac{1}{2\w_k}(\la|P_k|^2\ra-\w_k^2\la|Q_k|^2\ra)\neq
0,\label{akNk} \EEA since the property~(\ref{free}) is no longer
valid.

As we mentioned before, a complete set of new renormalized variables
$\at_k$ can be constructed, so that the strongly nonlinear system
can be viewed as a system of ``free'' waves in the sense of
vanishing correlations and the power spectrum, i.e., the new
variables $\at_k$ satisfy the properties of free waves given in
Eqs.~(\ref{asa}) and~(\ref{aa}). Next, we show how to construct
these renormalized variables $\at_k$.

Consider the generalization of the transformation~(\ref{intro_a}),
namely, the transformation from the Fourier variables $Q_k$ and
$P_k$ to the renormalized variables $\at_k$ by \BEA
\at_k=\frac{P_k-\i\wt_kQ_k}{\sqrt{2\wt_k}},\label{at} \EEA where
$\wt_k$ is an arbitrary function with the only restrictions \BEA
\wt_k>0,~~\wt_k=\wt_{N-k}.\label{canonical} \EEA One can show that,
these restrictions~(\ref{canonical}) provide a necessary and
sufficient condition for the transformation~(\ref{at}) to be
canonical. For the renormalized waves $\at_k$, we can compute \BEA
\la \at_k^*\at_l\ra&=&\frac{1}{2\wt_k}(\la|P_k|^2\ra+\wt_k^2\la|Q_k|^2\ra)\delta^k_l,\label{asat}\\
\la
\at_k\at_l\ra&=&\frac{1}{2\wt_k}(\la|P_k|^2\ra-\wt_k^2\la|Q_k|^2\ra)\delta^{k+l}_N.\label{aat}
\EEA Since we have the freedom of choosing any $\wt_k$ (with the
only restrictions~(\ref{canonical})), we can chose $\wt_k$ such that
$\la \at_k\at_{N-k}\ra$ vanishes. Thus, the renormalized variables
$\at_k$ for a strongly nonlinear system will behave like the bare
variables $a_k$ for a noninteracting system in terms of vanishing
correlations between waves. Therefore, we determine $\wt_k$ via \BEA
\la|P_k|^2\ra-\wt_k^2\la|Q_k|^2\ra=0.\label{vanish} \EEA Note that
the requirement~(\ref{vanish}) has the form of the virial theorem
for the free waves but with the renormalized linear dispersion
$\wt_k$. We rewrite Eq.~(\ref{vanish}) in terms of the kinetic and
quadratic potential parts of the energy of the mode $k$ as \BEA
\frac{\wt_k}{\w_k}=\sqrt{\frac{\la K_k\ra}{\la
U_k\ra}}.\label{vanish2} \EEA The  in Eqs.~(\ref{Kkind})
and~(\ref{Qkind}) leads to the $k$ independence of the right-hand
side of Eq.~(\ref{vanish2}). This allows us to define \textit{the
renormalization factor} $\eta$ for \textit{all} $k$'s by \BEA
\eta\equiv\frac{\wt_k}{\w_k}=\sqrt{\frac{\la K\ra}{\la
U\ra}}.\label{eta} \EEA for dispersion $\w_k$. Here
$K=\sum_{k=1}^{N-1}K_k$ and $U=\sum_{k=1}^{N-1}U_k$ are the kinetic
and the quadratic potential parts of the total energy of the
system~(\ref{Ham_PQ}), respectively. Note that the way of
constructing the renormalized variables $\at_k$ via the precise
requirement of vanishing correlations between waves yields the exact
expression for the renormalization factor, which is valid for all
wave numbers $k$ and any strength of nonlinearity. The independence
of $\eta$ of the wave number $k$ is a consequence of the Gibbs
measure. This $k$ independence phenomenon has been observed in
previous numerical experiments~\cite{our_prl,Alabiso}. We will
elaborate on this point in the results of the  numerical experiment
presented in Section~\ref{sect_FPU}.

The immediate consequence of the fact that $\eta$ is independent of
$k$ is that the power spectrum of the renormalized waves possesses
the precise Rayleigh-Jeans distribution, i.e., \BEA
\tilde{n}_k=\frac{\T}{\wt_k},\label{nt} \EEA from Eq.~(\ref{asat}),
where $\tilde{n}_k=\la |\at_k|^2\ra$. Combining Eqs.~(\ref{intro_a})
and~(\ref{at}), we find the relation between the ``bare'' waves
$a_k$ and the renormalized waves $\at_k$ to be \BEA
a_k=\frac{1}{2}\left(\sqrt{\eta}+\frac{1}{\sqrt{\eta}}\right)\at_k+\frac{1}{2}\left(\sqrt{\eta}-\frac{1}{\sqrt{\eta}}\right)\at_{N-k}.\label{a2at}
\EEA Using Eq.~(\ref{a2at}), we obtain the following form of the
power spectrum for the bare waves $a_k$ \BEA
n_k=\frac{1}{2}\left(1+\frac{1}{\eta^2}\right)\frac{\T}{\w_k},\label{nkbare}
\EEA which is a modified Rayleigh-Jeans distribution due to the
renormalization factor $(1+1/\eta^2)/2$. Naturally, if the
nonlinearity becomes weak, we have $\eta\rightarrow 1$, and,
therefore, all the variables and parameters with tildes reduce to
the corresponding ``bare'' quantities, in particular,
$\wt_k\rightarrow\w_k$, $\at_k\rightarrow a_k$,
$\tilde{n}_k\rightarrow n_k$. It is interesting to point out that,
even in a strongly nonlinear regime, the ``free-wave'' form of the
Rayleigh-Jeans distribution is satisfied
\textit{exactly}~[Eq.~(\ref{nt})] by the renormalized waves. Thus,
we have demonstrated that even in the presence of strong
nonlinearity, the system in thermal equilibrium can still be viewed
statistically as a system of ``free'' waves in the sense of
vanishing correlations between waves and the power spectrum.

Note that, in the derivation of the formula for the renormalization
factor [Eq.~(\ref{eta})], we only assumed the nearest-neighbor
interactions, i.e., the potential is the function of $q_j-q_{j+1}$.
One of the well-known examples of such a system is the $\beta$-FPU
chain, where only the forth order nonlinear term in $V$ is present.
In the remainder of the article, we will focus on the $\beta$-FPU to
illustrate the framework of the renormalized waves $\at_k$.
\section{Numerical study of the $\beta$-FPU chain}
\label{sect_FPU} Since its introduction in the early 1950s, the
study of the FPU lattice~\cite{FPU} has led to many great
discoveries in mathematics and physics, such as soliton
theory~\cite{Toda}. Being non-integrable, the FPU system also became
intertwined with the celebrated Kolmogorov-Arnold-Moser
theorem~\cite{Licht}. Here, we extend our results of the thermalized
$\beta$-FPU chain, which were briefly reported in~\cite{our_prl}.

The Hamiltonian of the $\beta$-FPU chain is of the form \BEA
H=\sum_{j=1}^N\frac{1}{2}p_j^2+\frac{1}{2}(q_j-q_{j+1})^2+\frac{\beta}{4}(q_j-q_{j+1})^4,\label{H_FPU}
\EEA where $\beta$ is a parameter that characterizes the strength of
nonlinearity.

The canonical equations of motion of the $\beta$-FPU chain are \BEA
\dot{q}_j&=&\frac{\partial H}{\partial p_j}=p_j,\label{dyn_pq_1}\\
\dot{p}_j&=&-\frac{\partial H}{\partial q_j}=(q_{j-1}-2q_j+q_{j+1})\nonumber\\
&+&\beta\big[(q_{j+1}-q_j)^3-(q_j-q_{j-1})^3\big].\nonumber \EEA To
investigate the dynamical manifestation of the renormalized
dispersion $\wt_k$ of $\at_k$, we numerically integrate
Eq.~(\ref{dyn_pq_1}). Since we study the thermal equilibrium
state~\cite{Ford,Alabiso_fpu,Lepri_fpu,Carati} of the $\beta$-FPU
chain, we use random initial conditions, i.e., $p_j$ and $q_j$ are
selected at random from the uniform distribution in the intervals
$(-p_{\rm{max}},p_{\rm{max}})$ and $(-q_{\rm{max}},q_{\rm{max}})$,
respectively, with the two constraints that (i) the total momentum
of the system is zero and (ii) the total energy of the system $E$ is
set to be a specified constant. We have verified that the results
discussed in the paper do not depend on details of the initial data.
Note that the behavior of $\beta$-FPU for fixed $N$ is fully
characterized by only one parameter $\beta E$~\cite{Poggy}. We use
the sixth order symplectic Yoshida algorithm~\cite{Yoshida} with the
time step $dt=0.01$, which ensures the conservation of the total
system energy up to the ninth significant digit for a runtime
$\tau=10^6$ time units. In order to confirm that the system has
reached the thermal equilibrium state~\cite{Parisi}, the value of
the energy localization~\cite{Cretegny} was monitored via
$L(t)\equiv{N\sum_{j=1}^{N}G_j^2}/{(\sum_{j=1}^{N}G_j)^2}$, where
$G_j$ is the energy of the $j$-th particle defined as \BEA
G_j&=&\frac{1}{2}p_j^2+\frac{1}{4}\big[(q_j-q_{j+1})^2+(q_{j-1}-q_j)^2\big]\nonumber\\
&+&\frac{\beta}{8}\big[(q_j-q_{j+1})^4+(q_{j-1}-q_j)^4\big]. \EEA If
the energy of the system is concentrated around one site, then
$L(t)=O(N)$. Whereas, if the energy is uniformly distributed along
the chain, then $L(t)=O(1)$. In our simulations, in thermal
equilibrium states, $L(t)$ is fluctuating in the range of $1$-$3$.
Since our simulation is of microcanonical ensemble, we have
monitored various statistics of the system to verify that the
thermal equilibrium state that is consistent with the Gibbs
distribution~(canonical ensemble) has been reached. Moreover, we
verified that, for $N$ as small as 32 and up to as large as 1024,
the equilibrium distribution in the thermalized state in our
microcanonical ensemble simulation is consistent with the Gibbs
measure. We compared the renormalization factor~(\ref{eta}) by
computing the values of $\la K\ra$ and $\la U\ra$ numerically and
theoretically using the Gibbs measure and found the discrepancy of
$\eta$ to be within $0.1\%$ for $\beta=1$ and the energy density
$E/N=0.5$ for $N$ from 32 to 1024.

We now address numerically how the renormalized linear dispersion
$\wt_k$ manifests itself in the dynamics of the $\beta$-FPU system.
We compute the spatiotemporal spectrum $|\af_k(\w)|^2$, where
$\af_k(\w)$ is the Fourier transform of $\at_k(t)$. (Note that, for
simplicity of notation, we drop a tilde in $\af_k$.)
\begin{figure}
\includegraphics[width=3in, height=2.5in]{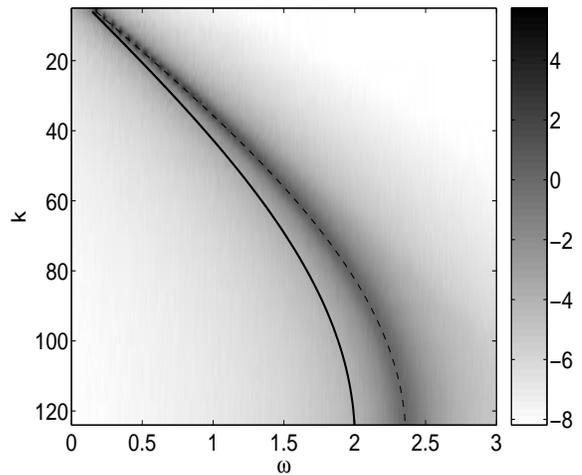}
\caption{The spatiotemporal spectrum $|\af_k(\w)|^2$ in thermal
equilibrium. The chain was modeled for $N=256$, $\beta=0.5$, and
$E=100$. [$\max\{-8,\ln{|\af_k(\w)|^2}\}$, with corresponding gray
scale, is plotted for a clear presentation]. The solid curve
corresponds to the usual linear dispersion $\w_k=2\sin(\pi k/N)$.
The dashed curve shows the locations of the actual frequency peaks
of $|\af_k(\w)|^2$.} \label{Fig_awk_num}
\end{figure}
Figure~\ref{Fig_awk_num} displays the spatiotemporal spectrum of
$\at_k$, obtained from the simulation of the $\beta$-FPU chain for
$N=256$, $\beta=0.5$, and $E=100$. In order to measure the value of
$\eta$ from the spatiotemporal spectrum, we use the following
procedure. For the fixed wave number $k$, the corresponding
renormalization factor $\eta(k)$ is determined by the location of
the center of the frequency spectrum $|\af_k(\w)|^2$, i.e., \BEA
\eta(k)=\frac{\w_{c}(k)}{\w_k},~~\mbox{with}~~
\w_{c}(k)=\frac{{\int}
\w|\af_k(\w)|^2~d\w}{\int|\af_k(\w)|^2~d\w}.\nonumber \EEA The
renormalization factor $\eta(k)$ of each wave mode $k$ is shown in
Fig.~\ref{Fig_et}~(inset). The numerical approximation $\bar{\eta}$
to the value of $\eta$ is obtained by averaging all $\eta(k)$, i.e.,
\BEA \bar{\eta}=\frac{1}{N-1}\sum_{k=1}^{N-1}\eta(k)\nonumber. \EEA
The renormalization factor for the case shown in
Fig.~\ref{Fig_awk_num} is measured to be $\bar{\eta}\approx1.1824$.
It can be clearly seen in Fig.~\ref{Fig_et}~(inset) that $\eta(k)$
is nearly independent of $k$ and its variations around $\bar{\eta}$
are less than $0.3\%$. We also compare the renormalization factor
$\eta$ obtained from Eq.~(\ref{eta}) (solid line in
Fig.~\ref{Fig_et}~(inset)) with its numerically computed
approximation $\bar{\eta}$ (dashed line in
Fig.~\ref{Fig_et}~(inset)). Equation~(\ref{eta}) gives the value
$\eta\approx 1.1812$ and the difference between $\eta$ and
$\bar{\eta}$ is less then $0.1\%$, which can be attributed to the
statistical errors in the numerical measurement.
\begin{figure}
\includegraphics[width=3in, height=2.5in]{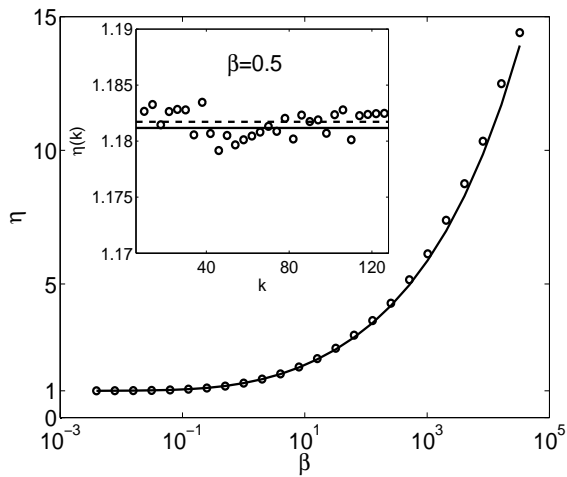}
\caption{The renormalization factor as a function of the
nonlinearity strength $\beta$. The analytical prediction
[Eq.~(\ref{eta})] is depicted with a solid line and the numerical
measurement is shown with circles. The chain was modeled for
$N=256$, and $E=100$. Inset: Independence of $k$ of the
renormalization factor $\eta(k)$. The circles correspond to
$\eta(k)$  obtained from the spatiotemporal spectrum shown in
Fig.~\ref{Fig_awk_num} [only even values of $k$ are shown for
clarity of presentation]. The dashed line corresponds to the mean
value $\bar{\eta}$. For $\beta=0.5$, the mean value of the
renormalization factor is found to be $\bar{\eta}\approx1.1824$. The
variations of $\eta_k$ around $\bar{\eta}$ are less then $0.3\%$.
[Note the scale of the ordinate.] The solid line corresponds to the
renormalization factor $\eta$ obtained from Eq.~(\ref{eta}). For the
given parameters $\eta\approx 1.1812$. } \label{Fig_et}
\end{figure}
In Fig.~\ref{Fig_et}, we plot the value of $\eta$ as a function of
$\beta$ for the system with $N=256$ particles and the total energy
$E=100$. The solid curve was obtained using Eq.~(\ref{eta}) while
the circles correspond to the value of $\eta$ determined via the
numerical spectrum $|\af_k(\w)|^2$ as discussed above. It can be
observed that there is excellent agreement between the theoretic
prediction and numerically measured values for a wide range of the
nonlinearity strength $\beta$.

In the following Sections, we will discuss how the renormalization
of the linear dispersion of the $\beta$-FPU chain in thermal
equilibrium can be explained from the wave resonance point of view.
In order to give a wave description of the $\beta$-FPU chain, we
rewrite the Hamiltonian~(\ref{H_FPU}) in terms of the renormalized
variables $\at_k$~[Eq.~(\ref{at})] with $\wt_k=\eta\w_k$, \BEA
H&=&\sum_{k=1}^{N-1}\frac{\w_k}{2}\left(\eta+\frac{1}{\eta}\right)|\at_k|^2\nonumber\\
&+&\frac{\w_k}{4}\left(\eta-\frac{1}{\eta}\right)(\at_k^*\at_{N-k}^*+\at_k\at_{N-k})\nonumber\\
&+&\sum_{k,l,m,s=1}^{N-1}T^{kl}_{ms}\Big[\D_{ms}^{kl}\at_k\at_l\at_m^*\at_s^*\nonumber\\
&+&\left(\frac{2}{3}\D^{klm}_{s}\at_k\at_l\at_m\at_s^*+c.c.\right)\nonumber\\
&+&\left(\frac{1}{6}\D^{klms}_{0}\at_k\at_l\at_m\at_s+c.c.\right)\Big],\label{H_a}
\EEA where c.c. stands for complex conjugate, and \BEA
T^{kl}_{ms}=\frac{3\beta}{8N\eta^2}\sqrt{\w_k\w_l\w_m\w_s}\label{T_inter}
\EEA is the interaction tensor coefficient. Note that, due to the
discrete nature of the system of finite size, the wave space is
periodic and, therefore, the ``momentum'' conservation is guaranteed
by the following ``periodic'' Kronecker delta functions \BEA
\D^{kl}_{ms}&\equiv&\delta_{ms}^{kl}-\delta^{klN}_{ms}-\delta^{kl}_{msN},\label{D22}\\
\D^{klm}_{s}&\equiv&\delta^{klm}_{s}-\delta^{klm}_{sN}+\delta^{klm}_{sNN},\label{D31}\\
\D^{klms}_{0}&\equiv&\delta^{klms}_{NN}-\delta^{klms}_{N}-\delta^{klms}_{NNN}.\label{D40}
\EEA Here, the Kronecker $\delta$-function is equal to 1, if the sum
of all superscripts is equal to the sum of all subscripts, and 0,
otherwise.
\section{Dispersion relation and resonances}
\label{sect_Dispersion} In order to address how the renormalized
dispersion arises from wave interactions, we study the resonance
structure of our nonlinear waves. Since the system~(\ref{H_a}) is a
Hamiltonian system with four-wave interactions, we will discuss the
properties of the resonance manifold associated with the $\beta$-FPU
system described by Eq.~(\ref{H_a}) as a first step towards the
understanding of its long time statistical behavior. We comment that
the resonance structure is one of the main objects of investigation
in wave turbulence
theory~\cite{ZLF,Newell,Lvov,Cai,MMT,Rink1,Rink2}. The theory of
wave turbulence focuses on the specific type of interactions, namely
resonant interactions, which dominate long time statistical
properties of the system. On the other hand, the non-resonant
interactions are usually shown to have a total vanishing average
contribution to a long time dynamics.

In analogy with quantum mechanics, where $a^+$ and $a$ are creation
and annihilation operators, we can view $\at_k^*$ as the
\textit{outgoing} wave with frequency $\wt_k$ and $\at_k$ as the
\textit{incoming} wave with frequency $\wt_k$. Then, the nonlinear
term $\at_k^*\at_l^*\at_m\at_s\Delta^{kl}_{ms}$ in
system~(\ref{H_a}) can be interpreted as the interaction process of
the type $(2\rightarrow 2)$, namely, two outgoing waves with wave
numbers $k$ and $l$ are ``created'' as a result of interaction of
the two incoming waves with wave numbers $m$ and $s$. Similarly,
$\at_k^*\at_l\at_m\at_s\Delta_{k}^{lms}$ in system~(\ref{H_a})
describes the interaction process of the type $(3\rightarrow 1)$,
that is, one outgoing wave with wave number $k$ is ``created'' as a
result of interaction of the three incoming waves with wave numbers
$l$, $m$, and $s$, respectively. Finally,
$\at_k\at_l\at_m\at_s\Delta_{0}^{klms}$ describes the interaction
process of the type $(4\rightarrow 0)$, i.e., all four incoming
waves interact and annihilate themselves. Furthermore, the complex
conjugate terms $\at_k\at_l^*\at_m^*\at_s^*\Delta_{lms}^{k}$ and
$\at_k^*\at_l^*\at_m^*\at_s^*\Delta^{0}_{klms}$ describe the
interaction processes of the type $(1\rightarrow 3)$ and
$(0\rightarrow 4)$, respectively.

Instead of the processes with the ``momentum'' conservation given
via the usual $\delta^{kl}_{ms}$, $\delta^{klm}_s$, or
$\delta^{klms}_0$ functions for an infinite discrete system, the
resonant processes of the $\beta$-FPU chain of a \textit{finite}
size are constrained to the manifold given by $\D^{kl}_{ms}$,
$\D^{klm}_s$, or $\D^{klms}_0$, respectively. Next, we describe
these resonant manifolds in detail. As will be pointed out in
Section~\ref{sect_Width}, there is a consequence of this
\textit{finite} size effect to the properties of the renormalized
waves.

The resonance manifold that corresponds to the $(2\rightarrow 2)$
resonant processes in the discrete periodic system, therefore, is
described by \BEA \BC
k+l\eqN m+s,\\
\wt_k+\wt_l=\wt_m+\wt_s, \EC\label{2to2} \EEA where we have
introduced the notation $g\eqN h$, which means that $g=h$, $g=h+N$,
or $g=h-N$ for any $g$ and $h$. The first equation in
system~(\ref{2to2}) is the ``momentum'' conservation condition in
the periodic wave number space. This ``momentum'' conservation comes
from $|\D^{kl}_{ms}|=1$. (Note that $|\D^{kl}_{ms}|$ can assume only
the value of $1$ or $0$.) Similarly, from $|\D^{klm}_{s}|=1$ and
$|\D^{klms}_{0}|=1$, the resonance manifolds corresponding to the
resonant processes of types $(3\rightarrow 1)$ and $(4\rightarrow
0)$ are given by \BEA \BC
k+l+m\eqN s,\\
\wt_k+\wt_l+\wt_m=\wt_s, \EC\label{3to1} \EEA and \BEA \BC
k+l+m+s\eqN 0,\\
\wt_k+\wt_l+\wt_m+\wt_s=0, \EC\label{4to0} \EEA respectively. For
the processes of type $(3\rightarrow 1)$, the notation $g\eqN h$
means that $g=h$, $g=h+N$, or $g=h+2N$. For the $(4\rightarrow 0)$
processes, $g\eqN h$ means that $g=h+N$, $g=h+2N$, or $g=h+3N$.

\begin{figure}
\includegraphics[width=2.5in, height=2.5in]{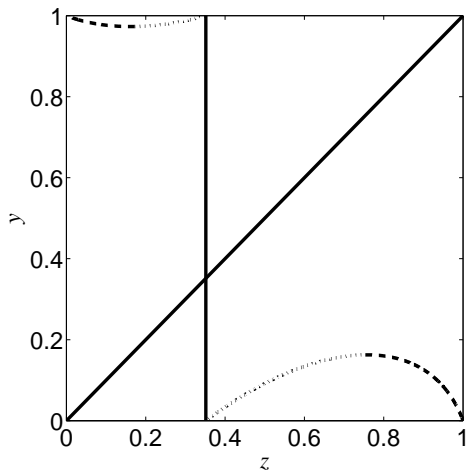}
\caption{The solutions of Eq.~(\ref{2to2cont}). The solid straight
lines correspond to the trivial resonances [solutions of
Eq.~(\ref{trivial})]. The solutions are shown for fixed $x=k/N$,
$k=90$, $N=256$ as the fourth wave number $v$ scans from ${1}/{N}$
to $(N-1)/{N}$ in the resonant quartet Eq.~(\ref{2to2cont}). The
non-trivial resonances are described by the dotted or dashed curves.
The dotted branch of the curves corresponds to the non-trivial
resonances described by Eq.~(\ref{folded1}) and and the dashed
branch corresponds to the non-trivial resonances described by
Eq.~(\ref{folded2}).} \label{Fig_resonances}
\end{figure}

\begin{figure}
\includegraphics[width=3in, height=2.5in]{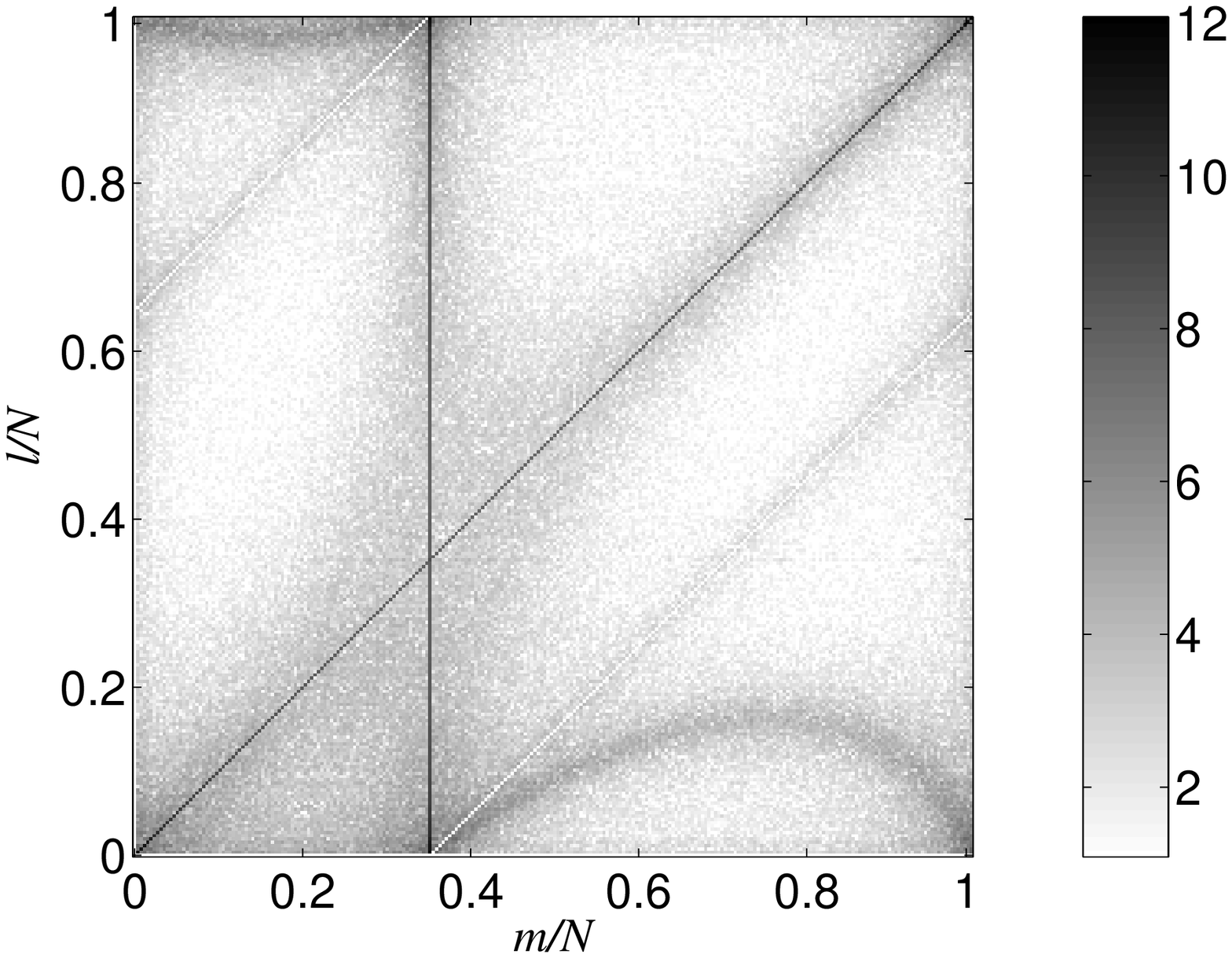}
\caption{The long time average
$|\la\at_k^*\at_l^*\at_m\at_s\ra\D^{kl}_{ms}|$ of the $\beta$-FPU
system in thermal equilibrium. The parameters for the FPU chain are
$N=256$, $\beta=0.5$, and $E=100$.
$\la\at_k^*\at_l^*\at_m\at_s\ra\D^{kl}_{ms}$ was computed for fixed
$k=90$. The darker grayscale corresponds to the larger value of
$\la\at_k^*\at_l^*\at_m\at_s\ra\D^{kl}_{ms}$. The exact solutions of
Eq.~(\ref{2to2cont}), which are shown in Fig.~\ref{Fig_resonances},
coincide with the locations of the peaks of
$|\la\at_k^*\at_l^*\at_m\at_s\ra\D^{kl}_{ms}|$. Therefore, the
darker areas represent the near-resonance structure of the finite
$\beta$-FPU chain. (The two white lines show the locations, where
$s=0$ and, therefore, $\at_k^*\at_l^*\at_m\at_s\D^{kl}_{ms}=0$.)
[$\max\{2,\ln(|\la\at_k^*\at_l^*\at_m\at_s\ra\D^{kl}_{ms}|)\}$ with
the corresponding grayscale is plotted for a clean presentation].}
\label{Fig_resonances_num}
\end{figure}

To solve system (\ref{2to2}), we rewrite it in a continuous form
with $x={k}/{N}$, $y={l}/{N}$, $z={m}/{N}$, $v={s}/{N}$, which are
real numbers in the interval $(0,1)$. By recalling that
$\wt_k=2\eta\sin({\pi k}/{N})$, we have \BEA \BC
x+y\eq1 z+v,\\
\sin(\pi x)+\sin(\pi y)=\sin(\pi z)+\sin(\pi v). \EC\label{2to2cont}
\EEA Thus, any rational quartet that satisfies Eq.~(\ref{2to2cont})
yields a solution for Eq.~(\ref{2to2}). There are two distinct types
of the solutions of Eq.~(\ref{2to2cont}). The first one corresponds
to the case \BEA x+y=z+v,\nonumber \EEA whose only solution is given
by \BEA \BC
x=z,\\
y=v, \EC \mbox{or}~~ \BC
x=v,\\
y=z,\label{trivial} \EC \EEA i.e., these are \textit{trivial}
resonances, as we mentioned above. The second type of the resonance
manifold of the $(2\rightarrow 2)$-type interaction processes
corresponds to \BEA x+y=z+v\pm 1,\nonumber \EEA the solution of
which can be described by the following two branches \BEA
z_1&=&\frac{x+y}{2}+\frac{1}{\pi}\arcsin(A)+2j,\label{folded1}\\
z_2&=&\frac{x+y}{2}-1-\frac{1}{\pi}\arcsin(A)+2j,\label{folded2}
\EEA where
$A\equiv\tan\left(\pi({x+y})/{2}\right)\cos\left(\pi({x-y})/{2}\right)$
and $j$ is an integer. The second type of resonances arises from the
discreteness of our model of a finite length, leading to
\textit{non-trivial} resonances. For our linear dispersion here,
non-trivial resonances are only those resonances that involve wave
numbers crossing the first Brillouin zone. As mentioned above, in
the setting of the phonon physics, these non-trivial resonant
processes are also known as the umklapp scattering processes. In
Fig.~\ref{Fig_resonances}, we plot the solution of
Eq.~(\ref{2to2cont}) for $x={k}/{N}$ with the wave number $k=90$ for
the system with $N=256$ particles (the values of $k$ and $N$ are
chosen merely for the purpose of illustration). We stress that
\textit{all} the solutions of the system~(\ref{2to2cont}) are given
by the Eqs.~(\ref{trivial}),~(\ref{folded1}), and~(\ref{folded2}),
and that the non-trivial solutions arise only as a consequence of
discreteness of the finite chain. The curves in
Fig.~\ref{Fig_resonances} represent the loci of $(z,y)$,
parametrized by the fourth wave number $v$, i.e., $x$, $y$, $z$, and
$v$ form a resonant quartet, where $z={m}/{N}$, and $y={l}/{N}$.
Note that the fourth wave number $v$ is specified by the
``momentum'' conservation, i.e., the first equation in
Eq.~(\ref{2to2cont}). The two straight lines in
Fig.~\ref{Fig_resonances} correspond to the trivial solutions, as
given by Eq.~(\ref{trivial}). The two curves (dotted and dashed)
depict the non-trivial resonances. Note that the dotted part of
non-trivial resonance curves corresponds to the
branch~(\ref{folded1}), and the dashed part corresponds to the
branch~(\ref{folded2}), respectively. An immediate question arises:
how do these resonant structures manifest themselves in the FPU
dynamics in the thermal equilibrium? By examining the
Hamiltonian~(\ref{H_a}), we notice that the resonance will control
the contribution of terms like
$\at_k^*\at_l^*\at_m\at_s\D^{kl}_{ms}$ in the long time limit.
Therefore, we address the effect of resonance by computing long time
average, i.e., $\la\at_k^*\at_l^*\at_m\at_s\ra\D^{kl}_{ms}$, and
comparing this average (Fig.~\ref{Fig_resonances_num}) with
Fig.~\ref{Fig_resonances}.
To obtain Fig.~\ref{Fig_resonances_num}, the $\beta$-FPU system was
simulated with the following parameters: $N=256$, $\beta=0.5$,
$E=100$, and the averaging time window $\tau=400\tilde{t}_1$, where
$\tilde{t}_1$ is the longest linear period, i.e.,
$\tilde{t}_1={2\pi}/{\wt_1}$. In Fig.~\ref{Fig_resonances_num}, mode
$k$ was fixed with $k=90$ and the mode $s$, a function of $k$, $l$,
and $m$, is obtained from the constraint $k+l\eqN m+s$, i.e.,
$|\D^{kl}_{ms}|=1$. Note that we do not impose here the condition
$\wt_k+\wt_l=\wt_m+\wt_s$, therefore,
$|\la\at_k^*\at_l^*\at_m\at_s\ra\D^{kl}_{ms}|$ is a function of $l$
and $m$. By comparing Figs.~\ref{Fig_resonances} and
\ref{Fig_resonances_num}, it can be observed that the locations of
the peaks of the long time average
$|\la\at_k^*\at_l^*\at_m\at_s\ra\D^{kl}_{ms}|$ coincide with the
loci of the $(2\rightarrow 2)$-type resonances. This observation
demonstrates that, indeed, there are nontrivial $(2\rightarrow
2)$-type resonances in the finite $\beta$-FPU chain in thermal
equilibrium. Furthermore, it can be observed in
Fig.~\ref{Fig_resonances_num} that, in addition to the fact that the
resonances manifest themselves as the locations of the peaks of
$|\la\at_k^*\at_l^*\at_m\at_s\ra\D^{kl}_{ms}|$, the structure of
\textit{near-resonances} is reflected in the finite width of the
peaks around the loci of the exact resonances. Note that, due to the
discrete nature of the finite $\beta$-FPU system, only those
solutions $x$, $y$, $z$, and $v$ of Eq.~(\ref{2to2cont}), for which
$Nx$, $Ny$, $Nz$, and $Nv$ are integers, yield solutions $k$, $l$,
$m$, and $s$ for Eq.~(\ref{2to2}). In general, the rigorous
treatment of the exact integer solutions of Eq.~(\ref{2to2}) is not
straightforward. For example, for $N=256$, we have the following two
exact quartets $\vec{k}=\{k,l,m,s\}$:
$\vec{k}=\{k,N/2-k,N/2+k,N-k\}$, $\vec{k}=\{k,N/2-k,N-k,N/2+k\}$ for
$k<N/2$, and $\vec{k}=\{k,3N/2-k,k-N/2,N-k\}$,
$\vec{k}=\{k,3N/2-k,N-k,k-N/2\}$ for $k>N/2$. We have verified
numerically that for $N=256$ there are no other exact integer
solutions of Eq.~(\ref{2to2}). In the analysis of the resonance
width in Section~\ref{sect_Width}, we will use the fact that the
number of exact non-trivial resonances [Eq.~(\ref{2to2})] is
significantly smaller than the total number of modes.

The broadening of the resonance peaks in
Fig.~\ref{Fig_resonances_num} suggests that, to capture the
near-resonances for characterizing long time statistical behavior of
the $\beta$-FPU system in thermal equilibrium, instead of
Eq.~(\ref{2to2}), one  needs to consider the following effective
system \BEA \BC
k+l\eqN m+s,\\
|\wt_k+\wt_l-\wt_m-\wt_s|<\Delta\w, \EC\label{2to2near} \EEA where
$0<\Delta\w\ll\wt_k$ for any $k$, and $\Delta\w$ characterizes the
resonance width, which results from the near-resonace structure.
Clearly, $\D\w$ is related to the broadening of the spectral peak of
each wave $\at_{\a}(t)$ with $\a=k,l,m$, or $s$ in the quartet, and
this broadening effect will be studied in detail in
Section~\ref{sect_Width}. Note that the structure of near-resonances
is a common characteristic of many periodic discrete nonlinear wave
systems~\cite{Lvov2,Colm, Pushkarev}.

Further, it is easy to show that the dispersion relation of the
$\beta$-FPU chain does not allow for the occurrence of
$(3\rightarrow 1)$-type resonances, i.e., there are no solutions for
Eq.~(\ref{3to1}), and, therefore, all the nonlinear terms
$\at_k^*\at_l\at_m\at_s\D^{k}_{lms}$ are non-resonant and their long
time average $\la\at_k^*\at_l\at_m\at_s\ra\D^{k}_{lms}$ vanishes. As
for the resonances of type $(4\rightarrow 0)$, since the dispersion
relation is non-negative, one can immediately conclude that the
solution of the system (\ref{4to0}) consists only of zero modes.
Therefore, the processes of type $(4\rightarrow 0)$ are also
non-resonant, giving rise to
$\la\at_k\at_l\at_m\at_s\ra\D_{0}^{klms}=0$. In this article, we
will neglect the higher order effects of the near-resonances of the
types $(3\rightarrow 1)$ and $(4\rightarrow 0)$.

In the following sections, we will study the effects of the resonant
terms of type $(2\rightarrow 2)$, namely, the linear dispersion
renormalization and the broadening of the frequency peaks of
$\at_k(t)$. It turns out that, the former is related to the trivial
resonance of type $(2\rightarrow 2)$ and the latter is related to
the near-resonances, as will be seen below.
\section{Self-consistency approach to frequency renormalization}
\label{sect_Renormalization} We now turn to the discussion of how
the trivial resonances give rise to the dispersion renormalization.
This question was examined in~\cite{our_prl} before. There, it was
shown that the renormalization of the linear dispersion of the
$\beta$-FPU chain arises due to the collective effect of the
nonlinearity. In particular, the trivial resonant interactions of
type $(2\rightarrow 2)$, i.e., the solutions of Eq.~(\ref{trivial}),
enhance the linear dispersion (the renormalized dispersion relation
takes the form $\wt_k=\eta\w_k$ with $\eta>1$), and effectively
weaken the nonlinear interactions. Here, we further address this
issue and present a self-consistency argument to arrive at an
approximation for the renormalization factor $\eta$. As it was
mentioned above, the contribution of the non-resonant terms have a
vanishing long time effect to the statistical properties of the
system, therefore, in our self-consistent approach, we ignore these
non-resonant terms. By removing the non-resonant terms and using the
canonical transformation \BEA \at_k=\frac{P_k-\i\eta_{sc}\w_k
Q_k}{\sqrt{2\eta_{sc}\w_k}},\label{a_sc} \EEA where $\eta_{sc}$ is a
factor to be determined, we arrive at a simplified effective
Hamiltonian from Eq.~(\ref{H_a}) for the finite $\beta$-FPU system
\BEA
H_{\rm{eff}}&=&\sum_{k=1}^{N-1}\frac{\w_k}{2}\left(\eta_{sc}+\frac{1}{\eta_{sc}}\right)|\at_k|^2\label{H_a_res}\\
&+&\sum_{k,l,m,s=1}^{N-1}T^{kl}_{ms}\Delta^{kl}_{ms}\at_k^*\at_l^*\at_m\at_s.\nonumber
\EEA The ``off-diagonal'' quadratic terms $\at_k\at_{N-k}$ from
Eq.~(\ref{H_a}) are not present in Eq.~(\ref{H_a_res}), since
$\at_k$ are chosen so that $\la\at_k\at_{N-k}\ra=0$ (see
Section~\ref{sect_Hformalism}). The contribution of the
\textit{trivial} resonances in $H_{\rm{eff}}$ is \BEA
H_4^{\rm{tr}}=4\sum_{k,l=1}^{N-1}T^{kl}_{kl}|\at_l|^2|\at_k|^2,\label{H4_diag}
\EEA which can be ``linearized'' in the sense that averaging the
coefficient in front of $|\at_k|^2$ in $H_4^{\rm{tr}}$ gives rise to
a quadratic form \BEA
H_2^{\rm{tr}}\equiv\sum_{k=1}^{N-1}\left(4\sum_{l=1}^{N-1}T^{kl}_{kl}\la|\at_l|^2\ra\right)|\at_k|^2.\nonumber
\EEA Note that the subscript $2$ in $H_2^{\rm{tr}}$ emphasizes the
fact that $H_2^{\rm{tr}}$ now can be viewed as a Hamiltonian for the
free waves with the familiar effective linear dispersion
$\Omega_k=4\sum_{l=1}^{N-1}T^{kl}_{kl}\la|\at_l|^2\ra$~\cite{our_prl,ZLF}.
This linearization is essentially a mean-field approximation, since
the long-time average of trivial resonances in Eq.~(\ref{H4_diag})
is approximated by the interaction of waves $\at_k$ with background
waves $\la |\at_l|\ra$. The self consistency condition, which
determines $\eta_{sc}$, can be imposed as follows: the quadratic
part of the Hamiltonian (\ref{H_a_res}), combined with the
``linearized'' quadratic part, $H_2^{\rm{tr}}$, of the quartic
$H_4^{\rm{tr}}$, should be equal to an effective quadratic
Hamiltonian $\tilde{H}_2=\sum_{k=1}^{N-1}\wt_k|\at_k|^2$ for the
renormalized waves, i.e., \BEA
& &\sum_{k=1}^{N-1}\frac{\w_k}{2}\left(\eta_{sc}+\frac{1}{\eta_{sc}}\right)|\at_k|^2\nonumber\\
&
&+\sum_{k=1}^{N-1}\left(4\sum_{l=1}^{N-1}T^{kl}_{kl}\la|\at_l|^2\ra\right)|\at_k|^2=\sum_{k=1}^{N-1}\wt_k|\at_k|^2,\nonumber\\\label{self}
\EEA where $\wt_k$ is the renormalized linear dispersion, which is
used in the definition of our renormalized wave, Eq.~(\ref{a_sc}),
and $\wt_k=\eta_{sc}\w_k$. Equating the coefficients of
$\w_k|\at_k|^2$ on both sides for every wave number $k$ yields \BEA
\frac{1}{2}\left(\eta_{sc}+\frac{1}{\eta_{sc}}\right)+4\sum_{l=1}^{N-1}\frac{3\beta}{8N\eta_{sc}^2}\w_l\langle
|\at_l|^2\rangle=\eta_{sc},\nonumber \EEA where use is made of
Eq.~(\ref{T_inter}). After algebraic simplification, we have the
following equation for $\eta_{sc}$ \BEA
\eta_{sc}^3-\eta_{sc}=\frac{3\beta}{N}\sum_{l=1}^{N-1}\w_l\langle|\at_l|^2\rangle.\label{eq_eta}
\EEA Using the property~(\ref{asat}) of the renormalized normal
variables $\at_k$, we find the following dependence of
$\la|\at_k|^2\ra$ on $\eta_{sc}$, \BEA
\la|\at_l|^2\ra=\frac{1}{2\eta_{sc}\w_l}\left(\la|P_l|^2\ra+\eta_{sc}^2\w_l^2\la|Q_l|^2\ra\right).\label{rhs}
\EEA Combining Eqs.~(\ref{eq_eta}) and (\ref{rhs}) leads to \BEA
\eta_{sc}^4-A\eta_{sc}^2-B=0,\label{eq_eta2} \EEA where \BEA
A&=&1+\frac{3\beta}{2N}\sum_{l=1}^{N-1}\w_l^2\langle|Q_l|^2\rangle=1+\frac{3\beta}{N}\la U\ra,\nonumber\\
B&=&\frac{3\beta}{2N}\sum_{l=1}^{N-1}\langle|P_l|^2\rangle=\frac{3\beta}{N}\la
K\ra.\nonumber \EEA The only physically relevant solution of
Eq.~(\ref{eq_eta2}) is \BEA
\eta_{sc}=\sqrt{\frac{A+\sqrt{A^2+4B}}{2}}.\label{new_eta} \EEA The
constants $A$ and $B$ can be easily derived using the Gibbs measure.

Next, we compare the renormalization factor $\eta$ [Eq.~(\ref{eta})]
with its approximation $\eta_{sc}$ [Eq.~(\ref{new_eta})] from the
self-consistency argument. In Appendix~\ref{app_FPU}, we study in
detail the behavior of both $\eta$ and $\eta_{sc}$ in the two
limiting cases, i.e., when nonlinearity is small ($\beta\rightarrow
0$ with fixed total energy $E$), and when nonlinearity is large
($\beta\rightarrow\infty$ with fixed total energy $E$). As is shown
in Appendix~\ref{app_FPU}, for the case of small nonlinearity, both
$\eta$ and $\eta_{sc}$ have the same asymptotic behavior in the
first order of the small parameter $\beta$, \BEA
\eta&=&1+\frac{3E}{2N}\beta+O(\beta^2),\label{small_eta}\\
\eta_{sc}&=&1+\frac{3E}{2N}\beta+O(\beta^2).\nonumber \EEA Moreover,
in the case of strong nonlinearity $\beta\rightarrow\infty$, both
$\eta$ and $\eta_{sc}$ scale as $\beta^\frac{1}{4}$, i.e., \BEA
\eta\sim\eta_{sc}\sim\beta^{\frac{1}{4}}\label{large_eta} \EEA (see
Appendix~\ref{app_FPU} for details). Note that, in~\cite{our_prl},
we numerically obtained the scaling $\eta\sim\beta^{0.2}$, which
differs from the exact analytical result~(\ref{large_eta}) due to
statistical errors in the numerical estimate of the power.
\begin{figure}
\includegraphics[width=3in, height=3in]{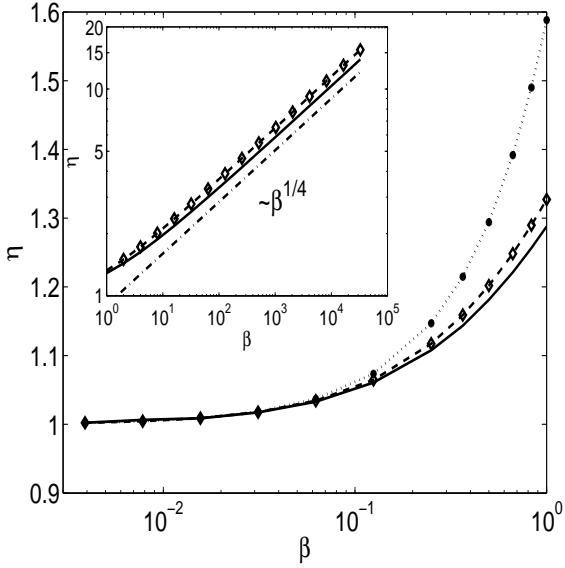}
\caption{The renormalization factor as a function of the
nonlinearity strength $\beta$ for small values of $\beta$. The
renormalization factor $\eta$ [Eq.~(\ref{eta})] is shown with the
solid line. The approximation $\eta_{sc}$ [Eq.~(\ref{new_eta})](via
the self-consistency argument) is depicted with diamonds connected
with the dashed line. The small-$\beta$ limit
[Eq.~(\ref{small_eta})] is shown with the solid circles connected
with the dotted line. Note that, abscissa is of logarithmic scale.
Inset: The renormalization factor as a function of the nonlinearity
strength $\beta$ for large values of $\beta$. The renormalization
factor $\eta$ [Eq.~(\ref{eta})] is shown with the solid line.
$\eta_{sc}$ [Eq.(\ref{new_eta})] is depicted with diamonds connected
with the dashed line. The large-$\beta$ scaling
[Eq.~(\ref{large_eta})] is shown with the dashed-dotted line. Note
that, the plot is of log-log scale with base 10.}
\label{Fig_eta_small}
\end{figure}
In Fig.~\ref{Fig_eta_small}, we plot the renormalization factor
$\eta$ and its approximation $\eta_{sc}$ for the case of small
nonlinearity $\beta$ for the system with $N=256$ particles and total
energy $E=100$. The solid line shows $\eta$ computed via
Eq.~(\ref{eta}), the diamonds with the dashed line represent the
approximation via Eq.~(\ref{new_eta}), and the solid circles with
the dotted line correspond to the small-$\beta$
limit~(\ref{small_eta}). In Fig.~\ref{Fig_eta_small}~(inset), we
plot the renormalization factor $\eta$ and its approximation
$\eta_{sc}$ for the case of large nonlinearity $\beta$ for the
system with $N=256$ particles and total energy $E=100$. The solid
line shows $\eta$ computed via Eq.~(\ref{eta}), the diamonds with
the dashed line represent the approximation via Eq.~(\ref{new_eta}),
and the dashed-dotted line correspond to the large-$\beta$
scaling~(\ref{large_eta}). Figure~\ref{Fig_eta_small} shows good
agreement between the renormalization factor $\eta$ and its
approximation $\eta_{sc}$ from the self consistency argument for a
wide range of nonlinearity, from $\beta\sim 10^{-3}$ to $\beta\sim
10^4$. This agreement demonstrates, that (i) the effect of the
linear dispersion renormalization, indeed, arises mainly from the
trivial four-wave resonant interactions, and (ii) our
self-consistency, mean-field argument is not restricted to small
nonlinearity.
\section{Resonance width}
\label{sect_Width} Finally, we address the question of how coherent
these renormalized waves are, i.e., we study how the nonlinear
interactions of waves in thermal equilibrium broaden the
renormalized dispersion. We will obtain an analytical formula for
the spatiotemporal spectrum $|\af_k(\w)|^2$ for the $\beta$-FPU
chain and compare the numerically measured width of the frequency
peaks with the predicted width.

In the Hamiltonian~(\ref{H_a_res}), the nonlinear terms
corresponding to the trivial resonances have been absorbed into the
quadratic part via the effective renormalized dispersion $\wt_k$.
Therefore, the new effective Hamiltonian is \BEA
\bar{H}=\sum_{k=1}^{N-1}\wt_k|\at_k|^2+\sum_{k,l,m,s=1}^{N-1}\Tt^{kl}_{ms}\Delta^{kl}_{ms}\at_k^*\at_l^*\at_m\at_s,\label{Hef}
\EEA where \BEA \BC
\Tt^{kl}_{ms}=T^{kl}_{ms}=\displaystyle{\frac{3\beta}{8N\eta^2}}\sqrt{\w_k\w_l\w_m\w_s},~k\neq m,~\mbox{and}~k\neq s\\
\Tt^{kl}_{ms}=0,~~\mbox{otherwise.} \EC\label{Tt} \EEA The new
interaction coefficient $\Tt^{kl}_{ms}$ ensures that the terms that
correspond to the interactions with trivial resonances are not
doubly counted in the Hamiltonian~(\ref{Hef}) and are removed from
the quartic interaction. This new interactions in the quartic terms
include the exact non-trivial resonant and non-trivial near-resonant
as well as non-resonant interactions of the $(2\rightarrow 2)$-type.

We change the variables to the interaction picture by defining the
corresponding variables $b_k$ via \BEA
b_k=\at_ke^{\i\wt_kt},\nonumber \EEA so that, the dynamics governed
by the Hamiltonian (\ref{Hef}) takes the familiar form \BEA \i\dot
b_k=2\sum_{l,m,s=1}^{N-1}\Tt^{kl}_{ms}\Delta^{kl}_{ms}b_l^*b_mb_se^{\i\wt^{kl}_{ms}
t},\label{b_dyn} \EEA where
$\wt^{kl}_{ms}=\wt_k+\wt_l-\wt_m-\wt_s$~\cite{Lvov}. Without loss of
generality, we consider only the case of $k<N/2$. As we have noted
before, only for a very small number of quartets does
$\wt^{kl}_{ms}$ vanish \textit{exactly}, i.e., $\wt^{kl}_{ms}=0$. We
separate the terms on the RHS of Eq.~(\ref{b_dyn}) into two kinds
--- the first kind with $\wt^{kl}_{ms}=0$ that corresponds to exact
non-trivial resonances, and the second kind that corresponds to
non-trivial near-resonances and non-resonances. Since, in the
summation, the first kind contains far fewer terms than the second
kind, and all the terms are of the same order of magnitude, we will
neglect the first kind in our analysis. Therefore, Eq.~(\ref{b_dyn})
becomes \BEA \i\dot b_k&=&2{\sum_{l,m,s=1}^{N-1}}^{\prime}
\Tt^{kl}_{ms}\Delta^{kl}_{ms}b_l^*b_mb_se^{\i\wt^{kl}_{ms}
t},\label{b_dyn2} \EEA where the prime denotes the summation that
neglects the exact non-trivial resonances.

\begin{figure}
\includegraphics[width=3in, height=2.5in]{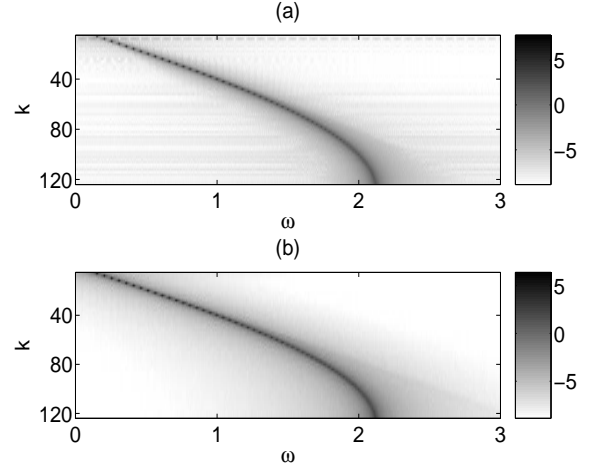}
\caption{ (a) Plot of the analytical prediction for the
spatiotemporal spectrum $|\af_k(\w)|^2$ via Eq.~(\ref{C}). (b) Plot
of the numerically measured spatiotemporal spectrum $|\af_k(\w)|^2$.
The parameters in both plots were $N=256$, $\beta=0.125$, $E=100$
and $\eta=1.06$, $\theta=0.401$. $\eta$ and  $\theta$ were computed
analytically via Gibbs measure. The darker gray scale correspond to
larger values of $|\af_k(\w)|^2$ in $\w$-$k$ space.
[$\max\{-8,\ln{|\af_k(\w)|^2}\}$ is plotted for clear
presentation].} \label{Fig_awk_both}
\end{figure}

\begin{figure}
\includegraphics[width=3in, height=2.5in]{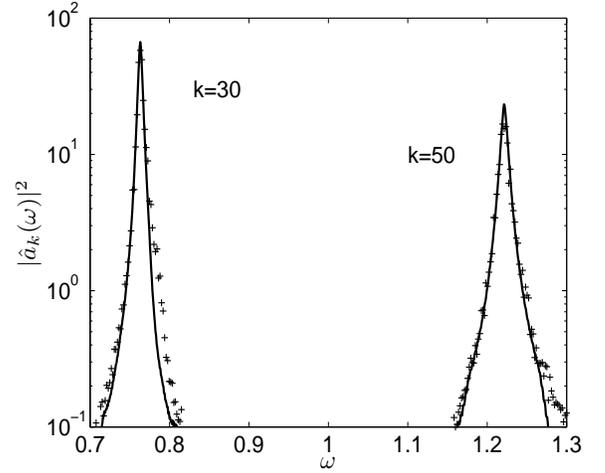}
\caption{Temporal frequency spectrum $|\af_k(\w)|^2$ for $k=30$
(left peak) and $k=50$ (right peak). The numerical spectrum is shown
with pluses and the analytical prediction [via Eq.~(\ref{C})] is
shown with solid line. The parameters were $N=256$, $\beta=0.125$,
$E=100$.
} \label{Fig_aw}
\end{figure}

\begin{figure}
\includegraphics[width=3in, height=2.5in]{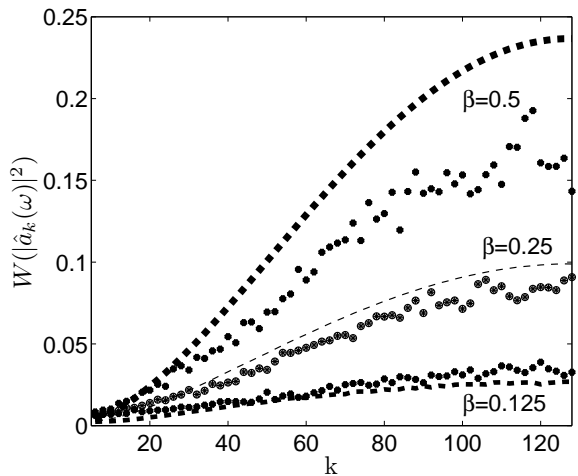}
\caption{Frequency peak width $W(|\af_k(\w)|^2)$ as a function of
the wave number $k$. The analytical prediction via Eq.~(\ref{C}) is
shown with a dashed line and the numerical observation is plotted
with solid circles. The parameters were $N=256$, $E=100$. The upper
thick lines correspond to $\beta=0.5$, the middle fine lines
correspond to $\beta=0.25$, and the lower solid circle and dashed
line (almost overlap) correspond to $\beta=0.125$.}
\label{Fig_widths}
\end{figure}

\begin{figure}
\includegraphics[width=3in, height=2.5in]{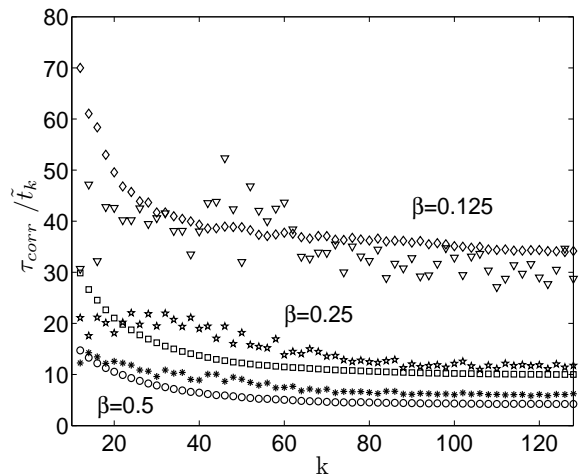}
\caption{ Ratio, as a function of $k$, of the correlation time
$\tau_k$ of the mode $k$ to the corresponding linear period
$\tt_k=2\pi/\wt_k$. Circles, squares, and diamonds represent the
analytical prediction for $\beta=0.5$, $\beta=0.25$, and
$\beta=0.125$ respectively. Stars, pentagrams, and triangles
correspond to the numerical observation for $\beta=0.5$,
$\beta=0.25$, and $\beta=0.125$ respectively. The parameters were
$N=256$, $E=100$. The ratio is sufficiently large for all wave
numbers $k$ even for relatively large $\beta=0.5$, which validates
the time-scale separation assumption used in deriving Eq.~(\ref{J}).
The comparison also suggests that for smaller $\beta$ the analytical
prediction should be closer to the numerical observation, as is
confirmed in Fig.~\ref{Fig_widths}.} \label{Fig_corr}
\end{figure}

The problem of broadening of spectral peaks now becomes the study of
the frequency spectrum of the dynamical variables $b_k(t)$ in
thermal equilibrium. This is equivalent to study the  two-point
correlation in time of $b_k(t)$ \BEA C_k(t)=\la b_k(t)b_k^*(0)\ra,
\EEA where the angular brackets denote the thermal average, since,
by Wiener-Khinchin theorem, the frequency spectrum \BEA
|b(\w)|^2=\mathfrak{F}^{-1}[C(t)](\w),\label{WKh} \EEA where
$\mathfrak{F}^{-1}$ is the inverse Fourier transform in time. Under
the dynamics (\ref{b_dyn2}), time derivative of the two-point
correlation becomes \BEA
\dot{C}_k(t)&=&\la \dot{b}_k(t)b_k^*(0)\ra\nonumber \\
&=&\la -2\i{\sum_{l,m,s}}^{\prime}\Tt^{kl}_{ms}b_l^*(t)b_m(t)b_s(t)e^{\i\wt^{kl}_{ms}t}\Delta^{kl}_{ms}b_k^*(0)\ra\nonumber\\
&=&-2\i{\sum_{l,m,s}}^{\prime}\Tt^{kl}_{ms}e^{\i\wt^{kl}_{ms}t}J^{kl}_{ms}(t)\Delta^{kl}_{ms}\label{cdot},
\EEA where \BEA J^{kl}_{ms}(t)\equiv\la
b_l^*(t)b_k^*(0)b_m(t)b_s(t)\ra.\nonumber \EEA In order to obtain a
closed equation for $C_k(t)$, we need to study the evolution of the
fourth order correlator $J^{kl}_{ms}(t)$. We utilize the weak
effective nonlinearity in Eq.~(\ref{Hef})~\cite{our_prl} as the
small parameter in the following perturbation analysis and obtain a
closure for $C_k(t)$, similar to the traditional way of deriving
kinetic equation, as in~\cite{ZLF,Benney}. We note that the
effective interactions of renormalized waves can be weak, as we have
shown in~\cite{our_prl}, even if the $\beta$-FPU chain is in a
strongly nonlinear regime. Our perturbation analysis is a multiple
time-scale, statistical averaging method. Under the near-Gaussian
assumption, which is applicable for the weakly nonlinear wave fields
in thermal equilibrium, for the four-point correlator, we obtain
\BEA
J^{kl}_{ms}(t)\Delta^{kl}_{ms}=C_k(t)C_l(0)(\delta^k_m\delta^l_s+\delta^k_s\delta^l_m).~~\label{4corr}
\EEA Combining Eqs.~(\ref{Tt}) and~(\ref{4corr}), we find that the
right-hand side of Eq.~(\ref{cdot}) vanishes because \BEA
\Tt^{kl}_{ms}J^{kl}_{ms}(t)\Delta^{kl}_{ms}=0. \EEA Therefore, we
need to proceed to the higher order contribution of
$J^{kl}_{ms}(t)$. Taking its time derivative yields \BEA
\dot{J}^{kl}_{ms}(t)\D^{kl}_{ms}&=&\la[\dot{b}_l^*(t)b_m(t)b_s(t)+b_l^*(t)\dot{b}_m(t)b_s(t)\nonumber\\
                &+&b_l^*(t)b_m(t)\dot{b}_s(t)]b_k^*(0)\ra\D^{kl}_{ms}.\label{Jdot}
\EEA Considering the right-hand side of Eq.~(\ref{Jdot}) term by
term, for the first term, we have \BEA
& &\la \dot{b}_l^*(t)b_m(t)b_s(t)b_k^*(0)\ra\D^{kl}_{ms}\nonumber\\
& &=\Bigg\la \Big[2\i{\sum_{\a,\b,\g}}^{\prime}\Tt^{l\a}_{\b\g}b_{\a}(t)b_{\b}^*(t)b_{\g}^*(t)\nonumber\\
& &\times
e^{-\i\w^{l\a}_{\b\g}}\D^{l\a}_{\b\g}\Big]b_m(t)b_s(t)b_k^*(0)\Bigg\ra\D^{kl}_{ms}.\label{4b}
\EEA We can use the near-Gaussian assumption to split the correlator
of the sixth order in Eq.~(\ref{4b}) into the product of three
correlators of the second order, namely, \BEA
& &\la b_k^*(0)b_m(t)b_s(t)b_{\a}(t)b_{\b}^*(t)b_{\g}^*\ra \D^{kl}_{ms}\nonumber\\
& &=C_k(t)n_m
n_s\delta^k_{\a}(\delta_{\b}^m\delta^s_{\g}+\delta^m_{\g}\delta^s_{\b}),\nonumber
\EEA where we have used that $n_m=C_m(0)$. Then, Eq.~(\ref{4b})
becomes \BEA
& &\la \dot{b}_l^*(t)b_m(t)b_s(t)b_k^*(0)\ra\D^{kl}_{ms}\nonumber\\
& &=4\i\Tt^{lk}_{ms}C_k(t)n_m n_s
e^{-\i\wt^{lk}_{ms}}\D^{kl}_{ms}.\label{4b1} \EEA Similarly, for the
remaining two terms in Eq.~(\ref{Jdot}), we have \BEA
& &\la b_l^*(t)\dot{b}_m(t)b_s(t)b_k^*(0)\ra\D^{kl}_{ms}\nonumber\\
& &=-4\i\Tt^{ms}_{kl}C_k(t)n_l n_s
e^{\i\wt^{ms}_{kl}}\D^{kl}_{ms},\label{4b2} \EEA and \BEA
& &\la b_l^*(t)b_m(t)\dot{b}_s(t)b_k^*(0)\ra\D^{kl}_{ms}\nonumber\\
& &=-4\i\Tt^{ms}_{kl}C_k(t)n_l n_m
e^{\i\wt^{ms}_{kl}}\D^{kl}_{ms},\label{4b3} \EEA respectively.
Combining Eqs.~(\ref{4b1}), (\ref{4b2}), and (\ref{4b3}) with
Eq.~(\ref{Jdot}), we obtain \BEA
\dot{J}^{kl}_{ms}(t)\D^{kl}_{ms}&=&4\i\Tt^{kl}_{ms}C_k(t)e^{-\i\wt^{kl}_{ms}t}\D^{kl}_{ms}\nonumber\\
& &\times(n_m n_s-n_l n_m-n_l n_s).\label{Jdot2} \EEA
Equation~(\ref{Jdot2}) can be solved for $J^{kl}_{ms}(t)$ under the
assumption that the term $e^{-\i\wt^{kl}_{ms}t}$ oscillates much
faster than $C_k(t)$. We numerically verify [Fig.~\ref{Fig_corr}
below] the validity of this assumption of time-scale separation.
Under this approximation, the solution of Eq.~(\ref{Jdot2})  becomes
\BEA
J^{kl}_{ms}(t)\D^{kl}_{ms}&=&4\Tt^{kl}_{ms}C_k(t)\D^{kl}_{ms}\frac{e^{-\i\wt^{kl}_{ms}t}-1}{-\wt^{kl}_{ms}}\nonumber\\
& &\times(n_m n_s-n_l n_m-n_l n_s )\label{J} \EEA Plugging
Eq.~(\ref{J}) into Eq.~(\ref{cdot}), we obtain the following
equation for $C_k(t)$ \BEA
\dot{C}_k(t)&=&8\i C_k(t) {\sum_{l,m,s}}^{\prime} \left(\Tt^{kl}_{ms}\right)^2\D^{kl}_{ms}\frac{1-e^{\i\w^{kl}_{ms}t}}{\w^{kl}_{ms}}\nonumber\\
& &\times(n_m n_s-n_l n_s-n_l n_m).\label{cdot2} \EEA Since in the
thermal equilibrium $n_k$ is known, i.e., $n_k=\la
|b_k(t)|^2\ra=\theta/\wt_k$ [Eq.~(\ref{nt})], Eq.~(\ref{cdot2})
becomes a closed equation for $C_k(t)$. The solution of
Eq.~(\ref{cdot2}) yields the autocorrelation function $C_k(t)$ \BEA
\ln\frac{C_k(t)}{C_k(0)}&=&8{\sum_{l,m,s}}^{\prime} \left(\Tt^{kl}_{ms}\right)^2\frac{e^{\i\w^{kl}_{ms}t}-1-\i\w^{kl}_{ms}t}{(\w^{kl}_{ms})^2}\nonumber\\
& &\times(n_l n_s+n_l n_m-n_m n_s)\D^{kl}_{ms}.\label{C} \EEA Using
this observation, together with Eq.~(\ref{Tt}), finally, we obtain
for the thermalized $\beta$-FPU chain \BEA
\ln\frac{C_k(t)}{C_k(0)}&=&\frac{9\beta^2\theta^2}{8N^2\eta^6}\w_k{\sum_{l,m,s}}^{\prime}(\w_m+\w_s-\w_l)\D^{kl}_{ms}\nonumber\\
&
&\times\frac{e^{\i\w^{kl}_{ms}t}-1-\i\w^{kl}_{ms}t}{(\w^{kl}_{ms})^2}.\label{C}
\EEA Equation~(\ref{C}) gives a direct way of computing the
correlation function of the renormalized waves $\at_k$, which, in
turn, allows us to predict the spatiotemporal spectrum
$|\af_k(\w)|^2$. In Fig.~\ref{Fig_awk_both}(a), we plot the
analytical prediction (via Eq.~(\ref{C})) of the spatiotemporal
spectrum
$|\af_k(\w)|^2\equiv|b_k(\w-\wt_k)|^2=\mathfrak{F}^{-1}[C(t)](\w-\wt_k)$.
By comparing this plot with the one presented in
Fig.~\ref{Fig_awk_both}(b), in which the corresponding numerically
measured spatiotemporal spectrum is shown, it can be seen that the
analytical prediction of the frequency spectrum via Eq.~(\ref{C}) is
in good qualitative agreement with the numerically measured one.
However, to obtain a more detailed comparison of the analytical
prediction with the numerical observation, we show, in
Fig.~\ref{Fig_aw}, the numerical frequency spectra of selected wave
modes with the corresponding analytical predictions. It can be
clearly observed that the agreement is rather good.
One of the important characteristics of the frequency spectrum is
the width of the spectrum. We compute the width $W(f)$ of the
spectrum $f(\w)$ by \BEA W(f)=\frac{\int f(\w)~d\w}{\max_\w f(\w)}.
\EEA In Fig.~\ref{Fig_widths}, we compare the width, as a function
of the wave number $k$, of the frequency peaks from the numerical
observation with that obtained from the analytical predictions. We
observe that, for weak nonlinearity ($\beta=0.125$), the analytical
prediction and the numerical observation are in excellent agreement.
In the weakly nonlinear regime, this agreement can be attributed to
the validity of (i) the near-Gaussian assumption, and (ii) the
separation between the linear dispersion time scale and the time
scale of the correlation $C_k(t)$. This separation was used in
deriving the analytical prediction [Eq.~(\ref{C})]. However, when
the nonlinearity becomes larger ($\beta=0.25$ and $\beta=0.5$), the
discrepancy between the numerical measurements and the analytical
prediction increases, as can be seen in Fig.~\ref{Fig_widths}.
Nevertheless, it is important to emphasize that, even for very
strong nonlinearity, our prediction is still qualitatively valid, as
seen in Fig.~\ref{Fig_widths}. In order to find out the effect of
the umklapp scattering due to the finite size of the chain, we also
computed the correlation [Eq.~(\ref{C})] with the ``conventional''
$\delta$-function $\delta^{kl}_{ms}$ (i.e., without taking into
account the umklapp processes) instead of our ``periodic'' delta
function $\D^{kl}_{ms}$. It turns out that the correlation time is
approximately $30\%$ larger if it is computed without umklapp
processes taken into account for the case $N=256$, $\beta=0.5$,
$E=100$. It demonstrates that the influence of the non-trivial
umklapp resonances is important and should be considered when one
describes the dynamics of the finite length chain of particles.
Finally, in Fig.~\ref{Fig_corr}, we verify the time scale separation
assumption used in our derivation, i.e., the correlation time of the
wave mode $k$ is sufficiently larger than the corresponding linear
dispersion period $\tt_k=2\pi/\wt_k$. In the case of small
nonlinearity ($\beta=0.125$), the two-point correlation changes over
much slower time scale than the corresponding linear oscillations
--- the correlation time is nearly two orders of magnitude larger than the corresponding linear oscillations for weak nonlinearity $\beta=0.125$, and
nearly one order of magnitude larger than the corresponding linear
oscillations for stronger nonlinearity $\beta=0.25$ and $\beta=0.5$.
This demonstrates that the renormalized waves have long lifetimes,
i.e., they are coherent over time-scales that are much longer than
their oscillation time-scales.
\section{Conclusions}
\label{sect_Conclusions} We have studied the statistical behavior of
the nonlinear periodic lattice with the nearest neighbor
interactions in thermal equilibrium. We have extended the notion of
normal modes to the nonlinear system by showing that regardless of
the strength of nonlinearity, the system in thermal equilibrium can
still be effectively characterized by a complete set of renormalized
waves, in the sense that those renormalized waves possess the
Rayleigh-Jeans distribution and vanishing correlations between
different wave modes. In addition, we have studied the property of
dispersion relation of the renormalized waves. The results we
obtained in Section~\ref{sect_Hformalism} are general and can be
applied to the large class of nonlinear systems with the nearest
neighbor interactions in thermal equilibrium.

We have further focused our attention on the particular system with
the nearest neighbor interactions --- the famous FPU chain. We have
confirmed that the general renormalization framework that we
discussed above is consistent with the numerical observations. In
particular, we have shown that the renormalized dispersion of the
thermalized $\beta$-FPU chain is in excellent agreement with the
numerical one for a wide range of the nonlinearity strength. We have
further demonstrated that the renormalized dispersion is a direct
consequence of the trivial resonant interactions of the renormalized
waves. Using a self-consistency argument, we have found an
approximation of the renormalization factor via a mean-field
approximation. In addition, we have used the multiple time-scale,
statistical averaging method to obtain the theoretical prediction of
the spatiotemporal spectrum and demonstrated that the renormalized
waves have long lifetimes. We note that the results obtained here
can be extended to general nonlinear potentials with the nearest
neighbor interactions.

The scenario of the wave behavior in the thermal equilibrium we
obtained here may suggest a theoretical framework for the
application of the wave turbulence to $\beta$-FPU in the situation
of \textit{near}-equilibrium.
\begin{acknowledgments}
We thank Sergey Nazarenko and Naoto Yokoyama for discussions. Y.L.
was supported by NSF CAREER DMS 0134955 and D.C. was supported by
NSF DMS 0507901.
\end{acknowledgments}
\appendix
\section{Limiting behaviors of $\eta$ for the thermalized $\beta$-FPU chain}
\label{app_FPU} We change variables $y_j=q_j-q_{j+1}$ in the
Hamiltonian~(\ref{H_FPU}) for $\beta$-FPU to obtain \BEA
H(p,y)=\sum_{j=1}^N\left[\frac{1}{2}p_j^2+\frac{1}{2}y_j^2+\frac{\beta}{4}y_j^4\right].\label{app2_Ham_py}
\EEA Next, we compute the pdf's for the momentum and displacement.
Any $p_j$ is distributed with the Gaussian pdf
$z_p=C_p\exp(-{\T}^{-1}p^2/2)$ and any $y_j$ is distributed with the
pdf $z_y=C_y\exp(-{\T}^{-1}(y^2+\beta y^4/2)/2)$, where $C_p$ and
$C_y$ are the normalizing constants. As we have discussed, the
renormalization factor $\eta$ of the $\beta$-FPU system in thermal
equilibrium is given by Eq.~(\ref{eta}), and its approximation via
the self-consistency argument $\eta_{sc}$ is given by
Eq.~(\ref{new_eta}). Here, we compare the behavior of both formulas
in two limiting cases, i.e., the case of small nonlinearity
$\beta\rightarrow 0$ and the case of strong nonlinearity
$\beta\rightarrow\infty$. We will use the following expressions for
the average density of kinetic, quadratic potential and quartic
potential parts of the total energy of the system \BEA
\frac{\la K\ra}{N}&=&\frac{1}{N}\sum_{j=1}^N\frac{\la p_j^2\ra}{2}=\frac{1}{2}\T,\label{app2_Hp2}\\
\frac{\la U\ra}{N}&=&\frac{1}{N}\sum_{j=1}^N\frac{\la y_j^2\ra}{2}=\frac{1}{2}\la y^2\ra,\label{app2_Hy2}\\
\frac{\la V\ra}{N}&=&\frac{\beta}{4N}\sum_{j=1}^N\la
y_j^4\ra=\frac{\beta}{4}\la y^4\ra.\label{app2_Hy4} \EEA In a
canonical ensemble, the temperature of a system is given by the
temperature of the heat bath. By identifying the average energy
density of the system with $\eb=E/N$ in our simulation (a
microcanonical ensemble), we can determine $\T$ as a function of
$\eb$ and $\beta$ by the following equation \BEA \frac{1}{N}\big(\la
K\ra+\la U\ra+\la V\ra\big)=\eb.\label{app2_eqT} \EEA We start with
the case of small nonlinearity $\beta\rightarrow 0$. Suppose in the
first order of the small parameter $\beta$ the temperature has the
following form \BEA \T(\beta)=\T_0+\beta \T_1\label{app2_Tsm}, \EEA
where $\T_0=O(1)$ and $\T_1=O(1)$. We find the values of $\T_0$ and
$\T_1$ using the constraint~(\ref{app2_eqT}). We use the following
expansions in the small parameter $\beta$ \BEA
& &\int_{-\infty}^{\infty}e^{-\frac{1}{2\T(\beta)}(y^2+\beta\frac{y^4}{2})}~dy\nonumber\\
&
&=\sqrt\frac{\pi}{8}\sqrt{\T_0}\left(4+\Big(\frac{2\T_1}{\T_0}-3\T_0\Big)\beta\right)+O(\beta^2),\label{app2_y0}
\EEA \BEA
& &\int_{-\infty}^{\infty}y^2e^{-\frac{1}{2\T(\beta)}(y^2+\beta\frac{y^4}{2})}~dy\nonumber\\
&
&=\sqrt\frac{\pi}{8}\sqrt{\T_0}\left(4\T_0+(6\T_1-15\T_0^2)\beta\right)+O(\beta^2),\label{app2_y2}
\EEA \BEA
& &\int_{-\infty}^{\infty}(y^2+\frac{\beta}{2}y^4)e^{-\frac{1}{2\T(\beta)}(y^2+\beta\frac{y^4}{2})}~dy\nonumber\\
&
&=\sqrt\frac{\pi}{8}\sqrt{\T_0}\left(4\T_0+(6\T_1-9\T_0^2)\beta\right)+O(\beta^2).\label{app2_y4}
\EEA Then, in the first order in $\beta$, Eq.~(\ref{app2_eqT})
becomes \BEA & &\T_0+\beta
\T_1+\frac{\sqrt\frac{\pi}{8}\sqrt{\T_0}\left(4\T_0+(6\T_1-9\T_0^2)\beta\right)}{\sqrt\frac{\pi}{8}\sqrt{\T_0}\left(4+
\Big(\frac{2\T_1}{\T_0}-3\T_0\Big)\beta\right)}=2\eb,\nonumber
\EEA and we obtain $\T_0=\eb$ and $\T_1=(3/4)\eb^2$. Therefore, for
the average kinetic energy density, we have \BEA \frac{\la
K\ra}{N}=\frac{1}{2}\eb+\frac{3}{8}\eb^2\beta+O(\beta^2),\label{app2_Kb}
\EEA and, for the average quadratic potential energy density, we
have \BEA \frac{\la
U\ra}{N}&=&\frac{1}{2}\frac{\sqrt\frac{\pi}{8}\sqrt{\T_0}\left(4\T_0+(6\T_1-15\T_0^2)\beta\right)}{\sqrt\frac{\pi}{8}\sqrt{\T_0}
\left(4+\Big(\frac{2\T_1}{\T_0}-3\T_0\Big)\beta\right)}\nonumber\\
&=&\frac{1}{2}\eb-\frac{9}{8}\eb^2\beta+O(\beta^2).\label{app2_U2b}
\EEA Finally, we obtain Eq.~(\ref{small_eta}), i.e., for small
$\beta$ \BEA
\eta=1+\frac{3}{2}\eb\beta+O(\beta^2).\label{app2_eta_approx} \EEA
Similarly, from Eq.~(\ref{new_eta}), we find the small $\beta$ limit
of the approximation $\eta_{sc}$ \BEA
\eta_{sc}=1+\frac{3}{2}\eb\beta+O(\beta^2).\nonumber \EEA Now, we
consider the case of strong nonlinearity $\beta\rightarrow\infty$.
From Eq.~(\ref{app2_eqT}), we conclude that temperature in the large
$\beta$ limit, which we denote as $\tinf$, stays bounded, i.e., $
0<\tinf<2\eb, $ and, in the limit of large $\beta$, we obtain for
Eq.~(\ref{app2_eqT}) \BEA
\tinf+\frac{\int_{-\infty}^{\infty}\frac{\beta}{2}y^4e^{-\frac{\beta}{4\tinf}y^4}~dy}{\int_{-\infty}^{\infty}e^{-\frac{\beta}{4\tinf}y^4}~dy}=2\eb.
\label{app2_eqT3} \EEA After performing the integration, we obtain $
\tinf=(4/3)\eb, $ and the average kinetic energy density becomes $
\la K\ra/N=(2/3)\eb $. For the average quadratic potential energy
density, we have \BEA
\frac{\la U\ra}{N}=\frac{1}{2}\frac{\int_{-\infty}^{\infty}y^2e^{-\frac{\beta}{4\tinf}y^4}~dy}{\int_{-\infty}^{\infty}e^{-\frac{\beta}{4\tinf}y^4}~dy}
=\frac{\Gamma(\frac{3}{4})}{\Gamma(\frac{1}{4})}\left(\frac{4\eb}{3\beta}\right)^{\frac{1}{2}}
\EEA For the renormalization factor, we obtain the following large
$\beta$ scaling \BEA
\eta=\sqrt{\frac{\Gamma(\frac{3}{4})}{\sqrt{3}\Gamma(\frac{1}{4})}}\eb^\frac{1}{4}\beta^\frac{1}{4}.\label{app2_etainf}
\EEA Similarly, for the approximation of $\eta_{sc}$, we obtain
$A=C\sqrt{\eb\beta}$, $B=4\eb\beta$, and
$C=2\sqrt{3}\Gamma(3/4)/\Gamma(1/4)$. Therefore, the large $\beta$
scaling of $\eta_{sc}$ becomes \BEA
\eta_{sc}=\sqrt{\frac{C+\sqrt{C^2+16}}{2}}\eb^\frac{1}{4}\beta^\frac{1}{4},\label{app2_etascinf}
\EEA which yields Eq.~(\ref{large_eta}).

\end{document}